\documentclass[11pt,preprint2]{aastex}

\begin{document}

\title{Discovery of thirteen novae candidates in M87\altaffilmark{1}}

%---------------------------------------------------------------------

\author{
Juan P. Madrid\altaffilmark{2}, 
William B. Sparks\altaffilmark{2},    
Henry C. Ferguson\altaffilmark{2},
Mario Livio\altaffilmark{2},
Duccio~Macchetto\altaffilmark{2,3}
}

\altaffiltext{1}{Based on  observations made with  the NASA/ESA Hubble
Space Telescope,  obtained at  the Space Telescope  Science Institute,
which is operated  by the Association of Universities  for Research in
Astronomy, Inc., under NASA  contract NAS 5-26555.  These observations
are associated with program 9474.}

\altaffiltext{2}{ Space  Telescope Science Institute,  3700 San Martin
Drive, Baltimore, MD 21218.}

\altaffiltext{3}{Affiliated with  the Space Telescope  Division of the
European   Space   Agency,    ESTEC,   Noordwijk,   The   Netherlands.}

%---------------------------------------------------------------------

\begin{abstract}

We  have made  thirteen positive  identifications  of near-ultraviolet
(NUV) transient sources  in the giant elliptical galaxy  M87 using the
Space Telescope Imaging Spectrograph  (STIS) on board the Hubble Space
Telescope (HST).  We give a  representative sample of the light curves
that   we  derive   for   these  transients,   and   based  on   their
characteristics  we identify  them as  classical novae  candidates. We
obtain a  hard lower limit for  the nova rate  in M87 of 64  novae per
year.  Our  results suggest  an enhancement on  the frequency  of nova
events towards  the nucleus  of the galaxy.   No correlation  is found
with  either jet  activity or  the  position of  present day  globular
clusters.

\end{abstract}


%---------------------------------------------------------------------

\section{Introduction} % Section 1

The  giant elliptical galaxy  M87, located  in the  core of  the Virgo
cluster, is well  known for featuring one of  the most spectacular and
best  studied extragalactic  jets.   M87's close  proximity, 16.1  Mpc
(Tonry  et al.   2001),  has  enabled HST  observations  at very  high
spatial resolution  of the  prominent jet, the  nucleus, and  the host
galaxy.   M87   has  thus  played   a  central  role  in   the  modern
understanding of jets and radio  loud active galaxies (see e.g. Junor,
Biretta, \& Livio 1999, and references therein).

During  the M87  jet  monitoring  program with  the  HST Faint  Object
Camera,  Sparks et  al.  (2000),  serendepitously discovered  at least
eleven transient  sources in the  host galaxy.  Based on  the temporal
variability and  their apparent luminosities, Livio,  Riess, \& Sparks
(2002), postulated  that these transients were  classical novae.  Even
though  Edwin Hubble reported  two probable  novae in  this particular
galaxy (Bowen, 1952,  Pritchet \& van den Bergh,  1987), until the era
of large ground based telescopes and HST, few novae have been observed
as far as Virgo (Ferrarese, et al.  1996, 2003).

Previous studies with HST have  found novae candidates near the center
of M87.   Shara et  al.  (2004) claimed  the discovery of  an erupting
classical nova in a globular cluster of M87 based on observations with
HST's Wide Field Planetary Camera 2.  Using STIS, Sohn et al.  (2006),
published a  catalogue of NUV-only  sources, where eleven of  them are
detected in only one epoch and are likely to be novae.

Here we report thirteen  NUV transient sources present in observations
made with STIS in 2003 and not found in the epochs analysed by Sohn et
al.   (2006). The transients  described here  are also  different from
those discovered by Sparks et al.  (2000). We show that these are most
likely classical novae close to the nucleus of M87.

%---------------------------------------------------------------------
\section{Observations} % Section 2

The STIS  observations of M87 that  we report were taken  under HST GO
program 9474 (PI: Sparks).  These observations were acquired within 51
days and were strategically divided  into three epochs to maximize the
likelihood of  finding new novae candidates and  to obtain information
on their  light curves.   The first observation  epoch starts  on 2003
June 7, the second epoch on 2003  July 12, and the third epoch on 2003
July 24.   Each epoch  comprises four visits  obtained one  day apart.
Each visit is divided into four exposures.  This amounts to a total of
512 minutes of STIS observations divided into 48 images of 640 seconds
each.

All  observations were  taken with  the NUV-MAMA  detector of  STIS; a
detailed description of  the instrument can be found  in Kimble et al.
(1998), and  Kim Quijano  et.  al. (2003).   The MAMA  detector yields
images  free of  cosmic  rays,  a critical  asset  when searching  for
transient sources with rapid  variability.  We use the ACCUM operating
mode  to obtain a  time-integrated image.   The filter  in use  is the
F25QTZ  with a  pivot wavelength  of 2364.8~\AA.   The detector  has a
field of  view of $24.7\arcsec\times24.7\arcsec$ and a  plate scale of
$\sim  0.024\arcsec$ per pixel.   At the  distance of  M87, $1\arcsec$
corresponds to 77 pc.

This project was aimed in  part at finding transient sources along the
jet.  The detector  covers the jet entirely but  provides only partial
coverage for the galaxy.  Jets were suspected of potentially enhancing
the rate  of classical novae  in their vicinity through  a jet-induced
accretion process (Livio, Riess,  \& Sparks, 2002).

%---------------------------------------------------------------------

\section{Data Reduction} % Section 3

We obtain the flatfielded  science files from the Multimission Archive
at Space Telescope (MAST).  These  files are calibrated using the code
version  2.17b of  the STScI  CALSTIS.  This  version of  the pipeline
applies the  geometrical distortion correction  for STIS.  We  use the
one-step pyraf  task Multidrizzle of the Space  Telescope Science Data
Analysis System (STSDAS) to combine  and rotate the images. The output
data has units of counts per second (Koekemoer et al.  2002).

Before generating  a catalogue of  NUV point sources  using SExtractor
(Bertin \& Arnouts, 1996) we  execute the following two steps to avoid
spurious detections. The  first step is to obtain  a galaxy subtracted
residual of each image.  For  this purpose we fit elliptical isophotes
to the STIS images using  the STSDAS task {\sc ellipse} (Jedrzejewski,
1987).  We construct a model of  M87 with the task {\sc bmodel} and we
subtract the model of the  galaxy from each exposure.  The second step
consists of masking the borders of the detector and the jet.

When  we run  SExtractor, the  parameters  best suited  to create  our
catalogue  of NUV  point sources  are: six  pixels above  threshold to
trigger detection and a detection  threshold of 2.2 $\sigma$.  We also
perform photometry  using SExtractor with an aperture  diameter of ten
pixels, i.e. $0.24\arcsec$.

The magnitudes  discussed throughout this paper are  determined in the
Space Telescope system derived with the expression:

\begin{displaymath}
m_{ST}  =  - 2.5log(\frac{PHOTFLAM  \times counts}{EXPTIME})-21.1
\end{displaymath} 

The {\sc photflam}  keyword is found on the header  of the STIS images
and it is equal to

\begin{mathletters}
{\sc photflam}=5.8455952 10$^{-18}$ erg s$^{-1}$ cm$^{-2}$ \AA$^{-1}$ [counts s$^{-1}$]$^{-1}$.
\end{mathletters}

Details  are  given  in the  HST  Data  Handbook  for STIS  (Brown  et
al. 2002).

We adopt the values of extinction used by Sohn et al.  (2006), namely,
$A_{NUV}  = 0.178$  mag,  and  an aperture  correction  of -0.230  mag
(Proffitt et al. 2003).

%---------------------------------------------------------------------

\section{Results} % Section 4

With  the  parameters we  set  for SExtractor  we  obtain  a total  of
thirteen  transient sources  that are  unequivocal  identifications. A
number of dubious detections  were eliminated by individual inspection
of the images.   With SExtractor we generate a  catalogue of NUV point
sources for each exposure allowing  us to derive the light curves.  We
also perform  photometry with SExtractor.  These point  sources span a
range of  5.3 magnitudes in  the STMAG system from  $m_{STIS}=22.3$ to
$m_{STIS}=27.6$, our  detection limit.  Within this  catalogue we find
point sources with roughly three  distinct types of light curves, only
one example of  each category is published given  the size constraints
of  this letter.  Table 1  gives the  position of  the  transients and
maximum magnitude they reach.

The first  category consists  of sources detected  in all  epochs with
constant brightness.   This category recovers the  sources of constant
brightness detected  by Sohn  et al.  (2006)  which are thought  to be
background galaxies.  In Figure 1 we plot the constant ``light curve''
of one  of these point sources:  NUV-04.  This light curve  is in good
agreement  with  the brightness  published  by  Sohn  et al.   (2006),
i.e. $m_{STIS} \approx 24.2$.   The sources of variable brightness are
of more interest for the present study and are described below.

The second category is composed  of the sources with overall declining
flux.   These sources are  present in  all, or  most, frames  and grow
fainter with  time.  Figure 2 shows  the light curve for  such a point
source.   This  source declines  slowly  at a  rate  of  about half  a
magnitude in 50 days.

A third category  consists of sources that flare  during the length of
our observations. The light curve of  such a source is shown in Figure
3. This  source  is  particularly  intriguing  because  of  its  rapid
variability. As this  source flared it was measured  in four different
frames at a magnitude of 23.8.   Within two days the source fell below
our detection limit and three days after flaring it has a magnitude of
25.1. This peculiar source was observed  towards the end of our survey
and consequently it only had a few days of coverage.

In  Figure  4  we  plot  the  positions  of   the  thirteen  positive
identifications of transient sources.  None of these sources appear in
other epochs, such as the ones  studied by Sohn et al.  (2006).  These
sources also have  light curves with variable brightness,  as shown in
Figures 2 \& 3.

\section{Discussion} 

\subsection{Are these transients novae?}

Are the variables we  detect classical novae indeed?  Previous studies
discuss  the  possible  different   origins  of  variable  sources  in
elliptical  galaxies of  the Virgo  cluster (Livio,  Riess  \& Sparks,
2002, Shara  et al.  2004  for M87; Ferrarese  et al.  2003  for M49).
All of  these studies  agree that the  variable sources they  find are
classical novae.  They  rule out other possible origins  such as GRBs,
or  background supernovae  given the  rarity of  these events  and the
small portion of the sky covered by HST detectors.

The  thirteen transients  we discovered  clearly belong  to  M87 given
their spatial distribution which overlaps the central starlight of the
galaxy  leaving the  western  section of  the  field of  view free  of
detections (see Figure 4).

The light  curves of these transients  show a slow rate  of decline in
the  NUV  consistent with  extensive  multiwavelength observations  of
other novae (e.g.  Nova Cygni 1978; Stickland et al.  1981). Generally
in  the  optical  the light  curve  of  a  nova declines  rapidly  and
monotonically, in the infrared it has large variations, while in the UV
after  a small  initial oscillation  the measured  flux  increases and
declines smoothly (see e.g. Warner (1995) for a discussion).

We identified these transient sources as novae based on their variable
brightness, their rate of decline, and the eruptive characteristics of
their light curves.

\subsection{Nova rate and spatial distribution}

Ciardullo  et al.  (1990)  and Shafter,  Ciardullo, \&  Prichet (2000)
define the nova rate $R$ for a  region of a galaxy using the mean nova
lifetime approach with the following relation:

\begin{displaymath}
R = \frac{N(m<m_{c})}{\tau_{c} +  \sum_{i=2}^n T_{i}}
\end{displaymath}

where

\begin{displaymath}
 T_{i} = min (t_{i}-t_{i-1}, \tau_{c}).
\end{displaymath}

In this relation $N(m<m_{c})$ is the number of observed novae brighter
than the critical  magnitude $m_{c}$.  $T_{i}$ is the  time sampled by
an individual frame taken at time  $t_{i }$.  $T_{i}$ is a function of
the  time  $t_{i-1}$  when  the  preceeding frame  was  taken  and  of
$\tau_{c}$, the length of time a nova remains brighter than $m_{c}$.

Given the  small number  of light  curves we were  able to  obtain, we
define an empirical, not a  modeled, mean nova lifetime for our survey
based on the length of time the thirteen detected novae were above our
detection threshold.  We find on  average $\tau_{c}$ = 28.07 days, and
$T_{i}  =  46.3$ days.   The  observed nova  rate  of  this survey  is
therefore

\begin{displaymath}
R=\frac{13}{74.4 \mathrm days}
\end{displaymath}

This is the equivalent of 64 novae per year.

This observed nova  rate is really a lower limit since  it needs to be
corrected  for the fraction  of novae  present in  the section  of the
galaxy not surveyed, novae lost  because of extinction, and novae with
brightness  below  our  detection   limit.   Table  2  summarizes  the
different nova rates for M87 found in the literature, and ranging from
70 to 300 events per year.

The  field of view  of STIS  covers $\sim  6 \%$  of the  total K-band
luminosity of  M87. Using the K-band  luminosity as a  scale to obtain
the global novae rate of the galaxy we would find the very large value
of 1067 events per year.

This  result suggests  that  the  frequency of  nova  events near  the
nucleus of M87 is considerably higher  than in the rest of the galaxy.
Crowding in the inner regions of M87 (Pritchet \& van den Bergh, 1987)
causes major incompleteness in the detection of novae within 25\arcsec
of the nucleus with ground-based data (Shafter et al. 2000) and within
5\arcsec of  the nucleus with  HST/WFPC2 data (Shara \&  Zurek, 2002).
Near-ultraviolet imaging  with HST,  allows detection of  novae within
1.5\arcsec of the nucleus of M87.

Moreover, 11 out  of 13 of our detections are  located within a radius
of $\sim  9 \arcsec$ of the nucleus.   It is at $\sim  9 \arcsec$ from
the  nucleus  that  the  light  distribution changes  from  a  shallow
power-law inside to a steep power-law outside (Lauer et al.  1992). In
Table 1  we give  the positions  of the transients  as the  offset and
radial  distance from  the center  of  M87. In  Figure 5  we show  the
cumulative   distribution   of  novae   events   and  the   cumulative
distribution  of V-band  (F555W)  light versus  radius.   As noted  by
Ferrarese  et al.   (2003) for  M49, the  novae are  more concentrated
towards the center than the galaxy light.

Our  findings indicate that  the novae  rates of  the inner  and outer
parts of  the galaxy  may be  different and one  should use  care when
scaling  to a  global  novae  rate using  the  K-band luminosity  (see
Williams \& Shafter (2004) for  a discussion).  An enhanced novae rate
towards  the  nucleus  may  be  due  to  massive  white  dwarfs  being
over-represented towards the center of  the galaxy as a result of mass
segregation.  Also, a release of  energy related to AGN activity might
play a role in enhancing the accretion rate of novae.  In fact, Sparks
et al. (1993) showed the presence of outflowing dust and gas filaments
near the nucleus of M87.

A comparison between the positions of the transients and the catalogue
of globular  clusters of  M87 made  by Kundu et  al.  (1999)  shows no
coincidence of their positions,  indicating that the transients do not
originate in present day globulars.

Unlike Livio, Riess, \& Sparks  (2002), who found an indication for an
enhanced rate of  novae in the vicinity of the  jet (see also Shafter,
Ciardullo \&Prichet  2000), we do not find  any particular orientation
or alignement  with respect to the  jet in the  thirteen transients we
detect. The  spatial distribution that we determine  matches well with
the  findings of  Sohn et  al. (2006)  and Shara  et al.   (2002), the
transient  sources described  by these  authors are  not conspicuously
oriented or aligned along the jet.

%---------------------------------------------------------------------

\acknowledgments

We are  grateful to David Floyd  and Leonardo Ubeda  (STScI) for their
help with writing an IDL code.  

%---------------------------------------------------------------------

%---------------------------------------------------------------------

%---------------------------------------------------------------------

\begin{deluxetable}{lcccc} 
\tablecaption{Transients positions and Maximum Magnitude\label{tbl-1}}
\tablewidth{0pt}     \tablehead{\colhead{ID}    &    \colhead{$\Delta$
R.A.\tablenotemark{a}}  & \colhead{$\Delta$  Decl.\tablenotemark{a}} &
\colhead{r\tablenotemark{b}} & \colhead{ Maximum Magnitude}}

\startdata

HST-uvnova1  & 1.13  & 1.03  & 1.52  & 23.84 $\pm$ 0.07 \\
HST-uvnova2  & 1.31  & 1.82  & 2.23  & 23.94 $\pm$ 0.08 \\
HST-uvnova3  & -2.29 & -0.20 & 2.27  & 24.13 $\pm$ 0.11 \\
HST-uvnova4  & 2.20  & 2.41  & 3.25  & 24.10 $\pm$ 0.09 \\
HST-uvnova5  & -0.35 & -3.76 & 3.77  & 23.97 $\pm$ 0.09 \\
HST-uvnova6  & -4.91 & -1.87 & 5.20  & 24.10 $\pm$ 0.09 \\
HST-uvnova7  & 5.58  & -2.23 & 5.96  & 23.08 $\pm$ 0.04 \\
HST-uvnova8  & -0.20 & 6.82  & 6.82  & 23.70 $\pm$ 0.07 \\
HST-uvnova9  & -5.79 & 4.47  & 7.26  & 24.34 $\pm$ 0.13 \\
HST-uvnova10 & -5.72 & 6.65  & 8.74  & 24.19 $\pm$ 0.11 \\
HST-uvnova11 & -8.94 & -1.57 & 9.00  & 23.78 $\pm$ 0.07 \\
HST-uvnova12 & -1.57 & 11.05 & 11.15 & 23.46 $\pm$ 0.12 \\
HST-uvnova13 & -8.12 & -8.04 & 11.37 & 23.78 $\pm$ 0.07 \\

 \enddata  

\tablenotetext{a}{Right ascension and declination offset in arcseconds
from the center of M87 at R.A. = $12^{h}30^{m}49.41^{s}$, decl = $12\arcdeg 23\arcmin28.96\arcsec$}

\tablenotetext{b}{Radial  distance  of  the  nova candidate  from  the
center of M87 in arcseconds. }

\end{deluxetable}

 %---------------------------------------------------------------------

\begin{deluxetable}{lcl} 
\tablecaption{M87 Nova Rates\label{tbl-2}}   \tablewidth{0pt}
\tablehead{\colhead{Reference}  & \colhead{$\eta$ yr$^{-1}$} & \colhead{Method}}

\startdata
Shafter et al. 2000   & 91 $\pm$ 34   & KPNO observations \\
Shara \& Zurek 2002   & $\gtrsim$ 300 & HST/WFPC2 observations\\
Matteucci et al. 2003 & 100-300       & Simulations\\
Ferrarese et al. 2003 & 70 $\pm$ 23   & Simulations\\

 \enddata  

\end{deluxetable}

%---------------------------------------------------------------------
\begin{figure}
\plotone{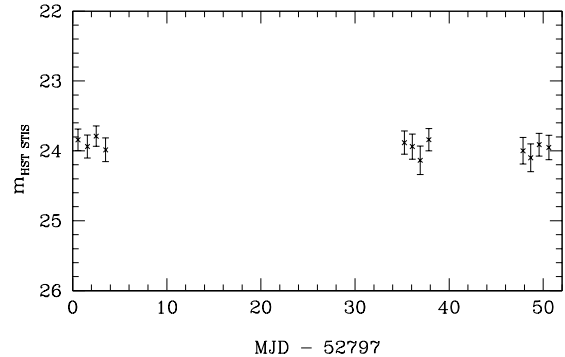}
\caption{Light  curve  of  NUV-04,  a source  of  constant  brightness
catalogued  by Sohn et al. 2006}
\end{figure} 

%---------------------------------------------------------------------
\begin{figure}
\plotone{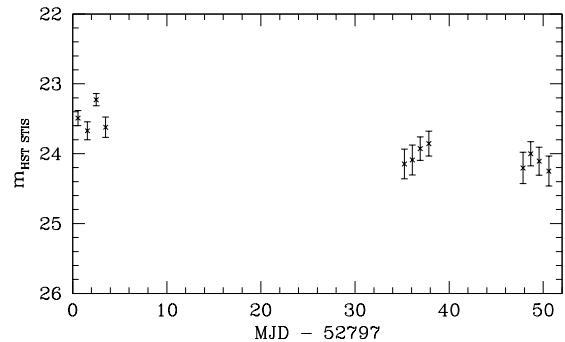}
\caption{Light curve of  HST-uvnova7, a transient with a  slow rate of
decline.}
\end{figure}

%---------------------------------------------------------------------
\begin{figure}
\plotone{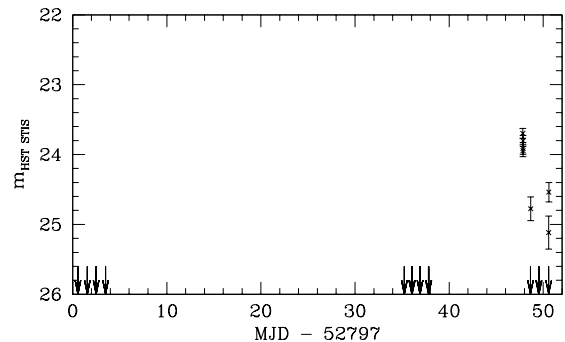}
\caption{Light curve  of HST-uvnova8, a transient  erupting during our
survey.}
\end{figure}

%---------------------------------------------------------------------
\begin{figure}
\plotone{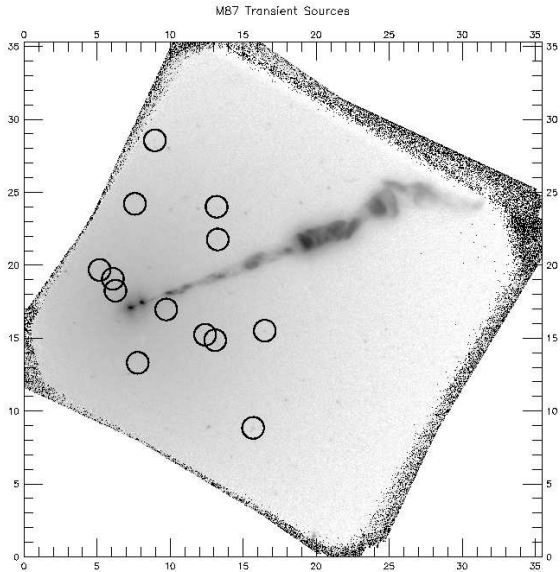}
\caption{HST/STIS NUV image of M87  showing the positions of the novae
candidates.}
\end{figure}

%---------------------------------------------------------------------

\begin{figure}
\plotone{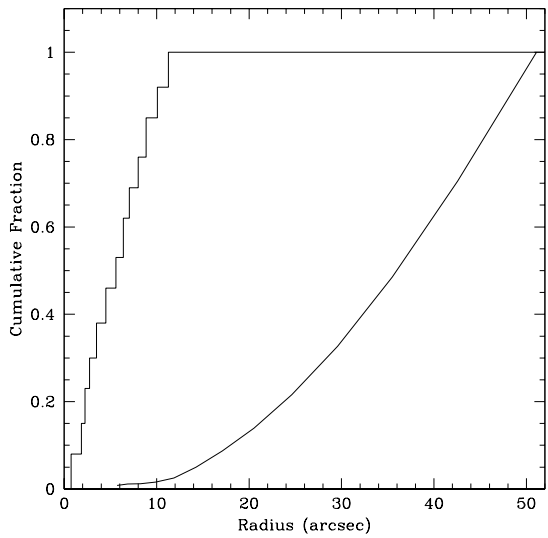}
\caption{Comparison between the cumulative distribution of nova events
(histogram) and the cumulative  fraction of V-band light (solid line).
The V-band light distribution was  obtained fitting {\sc ellipse} to a
WFPC2  F555W  exposure of  the  core of  M87.   The  novae are  highly
concentrated towards the center of the galaxy.}
\end{figure}
%---------------------------------------------------------------------
\end{document}